\documentclass[11pt,a4paper]{article}
\pdfoutput=1

\usepackage{epsfig}
\usepackage{latexsym}
\usepackage{amsfonts}
\usepackage{amsmath}
\usepackage{amsthm}
\usepackage{amssymb}
\usepackage{amsbsy}
\usepackage{multirow}
\usepackage{slashed}
\usepackage{color}
\usepackage{mathrsfs}
\usepackage{subfigure}
\usepackage[font={footnotesize,sl},bf]{caption}
\usepackage{wrapfig}

\usepackage{mathtools}
\usepackage{jheppub}
\usepackage{cancel}

\theoremstyle{definition} \newtheorem{result}{Result}
\theoremstyle{definition} \newtheorem{statement}{Statement}

\newcommand\be{\begin{equation}}
\newcommand\bea{\begin{eqnarray}}
\newcommand\ee{\end{equation}}
\newcommand\eea{\end{eqnarray}}

\newcommand{\bdm}{\begin{displaymath}}
\newcommand{\edm}{\end{displaymath}}
\newcommand{\nn}{\nonumber \\}
\newcommand{\f}[2]{\frac{#1}{#2}}
\newcommand{\bref}[1]{(\ref{#1})}
\newcommand\h{\frac{1}{2}}

\newcommand{\ket}[1]{|#1 \rangle}
\newcommand{\bra}[1]{\langle #1 |}

\newcommand{\defeq}{\vcentcolon=}

\usepackage{color}

\title{Cool horizons lead to information loss}
\author{Borun D. Chowdhury}
\affiliation{Department of Physics, \\
Arizona State University, \\
Tempe, Arizona 85287, USA
}

\abstract{There are two evidences for information loss during black hole evaporation:  (i) a pure state evolves to a mixed state and (ii) the map from the initial state to final state is non-invertible. Any proposed resolution of the information paradox must address both these issues. The firewall argument focuses only on the first and this leads to order one deviations from the Unruh vacuum for maximally entangled black holes. The nature of the argument does not extend to black holes in pure states.  It was shown by  Avery, Puhm and the author that requiring the initial state to final state map to be invertible mandates structure at the horizon even for pure states. The proof works if black holes can be formed in generic states and in this paper we show that this is indeed the case. We also demonstrate how models proposed by Susskind, Papadodimas et al. and  Maldacena et al. end up making the initial to final state map non-invertible and thus make the horizon ``cool" at the cost of unitarity.}


\begin{document}
\maketitle

\section{Introduction}

There are two results in Hawking's black hole evaporation analysis~\cite{Hawking:1974sw} which serve as evidence for loss of unitarity:
\begin{result} \label{Result1}
A pure initial state evolves to a mixed state,
\end{result}
\begin{result} \label{Result2}
The map from the initial state to the final state is many to one and thus non-invertible. 
\end{result}
Both of these originate from the assumption that the horizon of a black hole is in the Unruh vacuum, independent of the state of the matter that formed the black hole. Any proposed resolution of the information paradox must address/fix {\em both} these issues.\footnote{In the literature the term ``black hole" is used in two senses. In the strict sense it is a solution of classical gravity which leads to information loss and seems to have an infinite dimensional Hilbert space inside based on~\cite{Hawking:1974sw}. In a loose sense, it refers to the conjectured objects in quantum gravity which have a Hilbert space of size $e^{S_{BH}}$ and behave as black holes in as many ways as possible and yet  {\em somehow} evaporate unitarily. They are supposed to be the models one would get after fixing the issues listed above. Two such proposed models are the ``membranes" (of membrane paradigm and black hole complementarity)  and  the ``fuzzballs" (of the fuzzball proposal). We believe our usage will be clear from the context.} Note that in order to fix \hyperref[Result1]{Result 1}, the entropy of the outside radiation has to start decreasing no later than when half of the entropy of black hole has evaporated away~\cite{Page:1993df,Sen:1996ph}.

In Ref~\cite{Mathur:2009hf} Mathur demonstrated the stability of Hawking's result.  The essence of his proof is:
\begin{statement} \label{Statement1}
A pure state at the horizon implies increase in the entanglement entropy of radiation {\em at each step} of the evaporation. 
\end{statement}
In fact, this is only the leading order result  and Mathur actually showed that {\em small deviations from purity at the horizon} still imply an increase in the entanglement entropy of the radiation at each step of the evaporation. This statement implies \hyperref[Result1]{Result 1}. Since the Unruh vacuum is a pure state, this implies that small corrections to the Unruh vacuum at the horizon cannot fix \hyperref[Result1]{Result 1}. Mathur used this result to argue for structure at the horizon, a defining characteristic of fuzzballs. Recently Almheiri et. al. (AMPS)~\cite{Almheiri:2012rt} observed that the contrapositive of the above statement:
\begin{statement} \label{Statement2}
A pure final state implies a mixed state at the horizon {\em after half of the entropy of the black hole has evaporated (Page time)} 
\end{statement}
\noindent implies that, assuming purity of final state, a black hole maximally entangled with its radiation cannot have its horizon in the Unruh vacuum state. Claiming that  any other state would not allow  free infall, AMPS dubbed the state at the horizon a firewall.\footnote{It is not  obvious how one should see firewalls as distinct from fuzzballs.  We comment more on this in Section~\ref{StructurefulHorizonsOldResult}.}

The qualifier about when the horizon has to become a mixed state, comes from the result of Ref.~\cite{Page:1993df,Sen:1996ph} mentioned above. This has resulted in a discussion about ``young" (pure state) and ``old" (maximally entangled with radiation or some other system) black holes. While AMPS expressed a belief that firewalls must exist before Page time, the nature of the proof in~\cite{Mathur:2009hf,Almheiri:2012rt} (trying to address only \hyperref[Result1]{Result 1}) does not allow one to conclusively argue  that.\footnote{Note that \hyperref[Statement1]{Statement 1} of Mathur does not talk about young or old black holes. It simply states that a horizon in a pure state (or pure state up to small corrections) leads to increase in the entanglement of radiation. This of course precludes purity of the final state but does not treat black holes in pure or mixed states differently. We thank Samir Mathur for pointing this out.} 

In~\cite{Almheiri:2012rt} and most of the discussion following it, \hyperref[Result2]{Result 2}  has not been considered at all except in Ref.~\cite{Avery:2012tf} to the best of our knowledge. In~\cite{Avery:2012tf} the author, along with Avery and Puhm, took a different approach than Ref.~\cite{Mathur:2009hf}. We argued that purity of final state is necessary but not sufficient for unitarity. We showed that if one considers fixing \hyperref[Result2]{Result 2} also, i.e. demand that the initial to final state map be invertible, then there has to be an order one deviation from the Unruh vacuum at the horizon at every step of evaporation ( i.e. even before Page time/for pure states).\footnote{ Structure at the horizon for pure states was also predicted based on AdS/CFT in Ref.~\cite{Czech:2012be} before the firewall discussion started and was referred to as fuzzballs. Ref.~\cite{Bousso:1532382} argues for structure at the horizon for pure states which are Haar atypical but also expresses the opinion that such effects may cancel for Haar typical states. Ref.~\cite{Almheiri:2013hfa} has also argued for structure at the horizon for pure states based on different arguments about perturbative field theory modes in the bulk.}

Naively, there seems to be a caveat to our proof. We show that for generic states in the $e^{S_{BH}}$ sized Hilbert space of the black hole, the horizon cannot be in any fiducial state in general and the Unruh vacuum in particular. 
However, if the actual microstates of the black hole are special in that they live in a smaller dimensional sub Hilbert space then our result above does not apply for the following reason.  One may still conjecture some {\em special dynamics for special states} such that the horizon is predominantly in the Unruh vacuum before Page time~\cite{Susskind:2012uw}. In fact, allowing for some amount of non-locality (see for example~\cite{Papadodimas:2012aq}), {\em special non-local dynamics for special states} may allow the horizon to be predominantly in the Unruh vacuum at all times. One reason to suspect such special states for black holes would be that  't Hooft~\cite{tHooft:1993gx} showed that the information content of black holes formed from stellar collapse is $S_{BH}^{3/4}$ which is parametrically smaller that $S_{BH}$.

However, properties of black holes should not depend on how they are formed and whether or not they are in special states {\em unless} there is a fundamental obstruction to making black holes in non-special states. In this paper we show that it is possible to adiabatically collapse radiation in an arbitrary state living in a Hilbert space of size $e^{S_{BH}}$, to form a black holes in {\em any} of the $e^{S_{BH}}$ states. While this claim was made by Zurek in~\cite{PhysRevLett.49.1683}, it turns out to not work for Schwarzschild black holes. Nevertheless, it does work for the D1-D5-P black string. The difference comes from the sign of the specific heat in the two cases.
While we demonstrate the possibility of adiabatic collapse explicitly for the D1-D5-P black string, we expect it to work for any big black hole in AdS. After all, in the context of AdS/CFT the claim just amounts to saying we can excite the CFT into any one of the microscopic states accounting for the microcanonical entropy. There doesn't seem to be any fundamental obstruction to it. 

Having established this, our proof of~\cite{Avery:2012tf} is on a firm footing. We then analyze some recent attempts to do away with fuzzballs/firewalls~\cite{Susskind:2012uw,Papadodimas:2012aq,Maldacena:2013xja}. We find that the common theme in all of them is to find  a way to oppose \hyperref[Statement2]{Statement 2}. This is done in various ways (e.g. the so called $A=C$ models postulate the inside and the radiation are not independent systems). However, \hyperref[Statement2]{Statement 2} is closely related to \hyperref[Result1]{Result 1}. Whatever model one proposes, \hyperref[Result2]{Result 2} still needs to be taken care of to ensure unitarity.

Concretely, just by establishing that the ``horizon" (however a given model defines it) is pure does not mean it is in the Unruh vacuum. We find that one then has two ways to interpret these models:
\begin{itemize}
\item These models  are of the kind ``special dynamics for special states" with the fine tuned dynamics ensuring the horizon is always in the Unruh vacuum for special states. If we apply these to more generic states then the horizon will not be smooth.
\item One can understand the models as applying to all states. In such a case, we demonstrate that \hyperref[Result2]{Result 2} still stands and we end up loosing unitarity again as a cost for smooth horizons.
\end{itemize}

Thus the main result of the paper is that a smooth, information free horizon in a fiducial state (Unruh vacuum) generically leads to information loss. Thus, we conclude that if unitarity is to be preserved then the horizon of a black hole cannot be in the Unruh vacuum at any time during the evolution for generic states. Since a horizon cannot exist but in an Unruh vacuum state, our result shows that, as far as one can trust qubit models, quantum gravity microstates have to end in a (possibly complicated Planck scale) cap outside the horizon i.e. fuzzballs.

We go on to  discuss  how analytic continuation and  thermofield doubling are related and why one should not trust the results they provide for ``behind the horizon physics". We also propose a solution to the puzzle raised by Sen that if quantum gravity effects become important at horizon scale, why does semi-classical gravity work so well for black hole microstate counting.

While this paper was being written, Ref.~\cite{Marolf:2013dba}  discussing structure at horizons for pure states using different arguments appeared on the arXiv.  It also criticizes the claims of~\cite{Papadodimas:2012aq,Maldacena:2013xja} based on arguments independent of those in this paper.

The plan of the paper is as follows. In Section \ref{InformationLoss} we briefly review the information loss paradox, the stability of Hawking's result~\cite{Mathur:2009hf} and the firewall argument~\cite{Almheiri:2012rt}. In Section~\ref{BHStorage} we discuss how black holes may be formed in any of the $e^{S_{BH}}$ states by adiabatically compressing radiation in a box. In Section~\ref{StructurefulHorizons} we give a intuitive explanation of the proof of~\cite{Avery:2012tf} and explain how certain models proposed to do away with fuzzballs/firewalls either do not work generically, or lead to loss of unitarity depending on the interpretation. We end with conclusions and discussions in Section~\ref{Discussion}.

\section{Information paradox and small corrections} \label{InformationLoss}

\subsection{Leading order Hawking result}

In~\cite{Hawking:1974sw} Hawking showed that a black hole formed from the collapse of a pure state evaporates away into radiation which is in a thermal state. The final radiation being in a mixed state and the map from initial to final state being non-invertible both imply a breakdown of unitarity, a problem otherwise known as the information paradox. 

Hawking's process can be understood in terms of stretching of ``nice slices" (see~\cite{Lowe:1995ac}). When such slices in the vacuum state (Unruh vacuum) stretch, the stretching produces particle pairs at  intervals ($t \sim M$) which were shown in~\cite{Giddings:1992ff} to be given by
\be
\ket{\text{Unruh Vacuum}}=C e^{\gamma \hat b^\dagger b^\dagger} \ket{\hat 0} \ket{0}\,.
\ee
We have adopted the notation where hatted quantities are for the Hilbert space inside the black hole and unhatted ones are for the Hilbert space outside. $C$ is a normalization constant and $\gamma$ is an order one number. One can truncate this to the first two terms
\be
\ket{\varphi_1}\defeq \ket{\text{Unruh Vacuum}}=\f{1}{\sqrt 2} (\ket{\hat 0} \ket{0} + \ket{\hat 1} \ket{1}) \label{QubitUnruhVacuum}
\ee
to make the process analyzable using qubits. We view the black hole as a system and the outside as another system and the horizon as the interface of these two systems. 

In these terms, Hawking's result implies that a state of the black hole described by $n$ qubits 
\be
\ket{BH} = \sum C_{\hat q_1  \dots \hat q_n} \ket{\hat q_1  \dots \hat q_n }
\ee
evolves to
\bea
\ket{BH} 
&& \longrightarrow\ket{BH} \otimes \Big[ \f{1}{\sqrt 2} (\ket{\hat 0} \ket{0} + \ket{\hat 1} \ket{1}) \Big] \nn
&& \longrightarrow \ket{BH} \otimes \Big[ \f{1}{\sqrt 2} (\ket{\hat 0} \ket{0} + \ket{\hat 1} \ket{1}) \Big]^2 \nn
&& \vdots  \nn
&& \longrightarrow \ket{BH} \otimes \Big[ \f{1}{\sqrt 2} (\ket{\hat 0} \ket{0} + \ket{\hat 1} \ket{1}) \Big]^m \nn
&&\vdots  \label{HawkingEvolution}
\eea

This process terminates after $n$ steps because of energy conservation. Each emission carries away some energy leading to the black hole eventually evaporating away.\footnote{\label{numberOfEmissions}For Schwarzschild black holes, each emission carries away energy $(GM)^{-1}$ so the process stops after $GM^2$ steps. The entropy of the black hole is $e^{GM^2}$. So the map to qubit models is $GM^2 \sim n$.}
It is easy to see that the von Neumann entropy of the radiation outside starts at zero and grows by $\log 2$ at each step going upto $n \log 2$. Since we started with a pure state, {\em one} of the requirements for unitarity is that the entropy goes back to zero after $n$ steps.

However, one may think that this is a leading order process and there could be small corrections to it. These could originate from non-perturbative effects, for instance, making them  of the order of $e^{-GM^2}$.  The large number of states $e^{GM^2}$ could offset this smallness and restore unitarity. In the qubit models  the number of states is $2^n$ and so the small corrections would then be $\sim 2^{-n}$.

\begin{figure}[htbp]
\begin{center}
\subfigure[]{
\includegraphics[scale=.5]{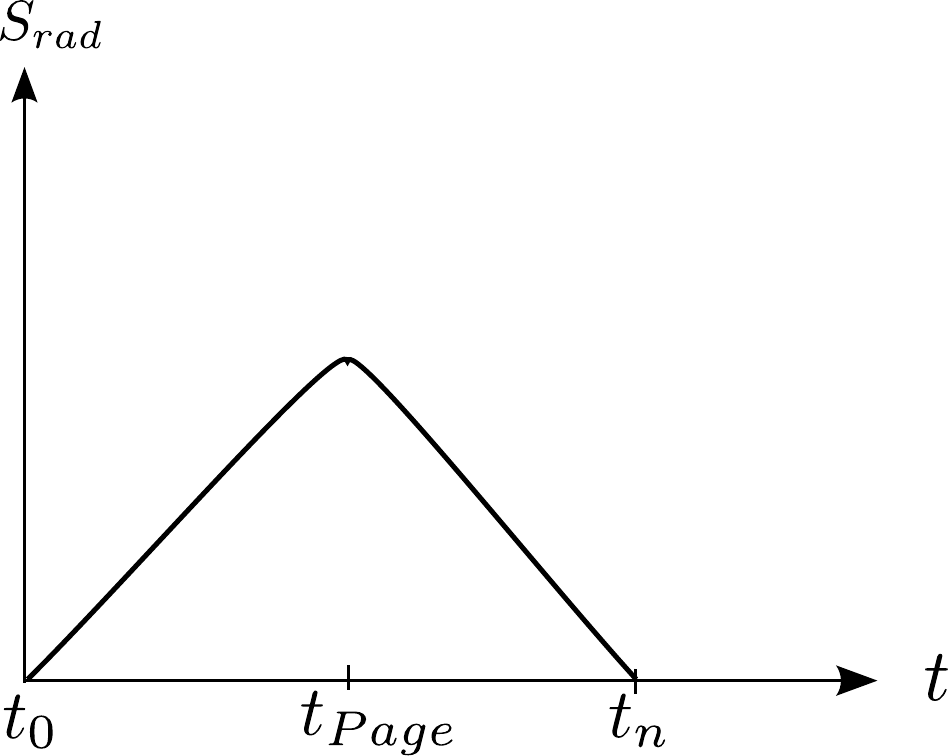}} \hspace{1in}
\subfigure[]{ 
\includegraphics[scale=.5]{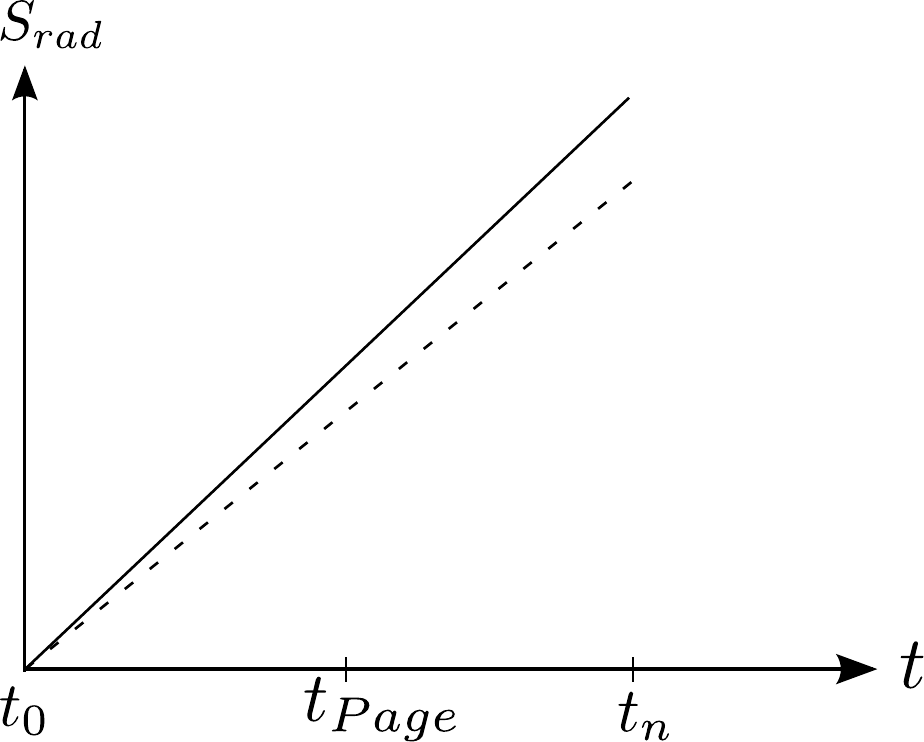} } 
\caption{
Entanglement entropy of radiation from  (a) a normal body in a typical state and (b) a traditional black hole with information-free horizon. In normal bodies in a typical state, the entropy initially goes up and then goes down while for traditional black holes that evaporate via Hawking-pair creation the entropy monotonically increases. Allowing small correction to the leading order process (solid line) decreases the slope (dashed line) but the entropy curve keeps rising. 
}
\label{evaporation}
\end{center}
\end{figure}

Page conjectured in~\cite{Page:1993df} (and this was proven in~\cite{Sen:1996ph}) that for typical states in a Hilbert space $\mathcal H$ of dimension $D_{\mathcal H}$, the von-Neuman entropy of a subsystem $\mathcal A$ of dimension $D_{\mathcal A}$ increases from zero with the dimension of $D_{\mathcal A}$    till $D_{\mathcal A}^2 = D_{\mathcal H}$ and then starts to decrease, going back to zero when $D_{\mathcal A} = D_{\mathcal H}$. So any small corrections would have to make the entropy of the radiation turn around after half the black hole has evaporated ($n/2$ steps) and bring it down to zero in the end ($n$ steps) as shown in  Figure~\ref{evaporation}a. This would be {\em one} of the requirements for unitarity.

\subsection{Small corrections to Hawking's process and purity of final state}

Note that the state at the interface is a two qubit system which, in principle, could be in a four dimensional Hilbert space spanned by the Bell states
\begin{equation}\begin{aligned}
\ket{\varphi_1} &\defeq \frac{1}{\sqrt{2}}\big(\ket{\hat{0} }\ket{0 } 
                           + \ket{\hat{1} }\ket{1 }\big)\,,\\
\ket{\varphi_2} &\defeq \frac{1}{\sqrt{2}}\big(\ket{\hat{0} }\ket{0 } 
                           - \ket{\hat{1} }\ket{1 }\big)\,,\\
\ket{\varphi_3} &\defeq \frac{1}{\sqrt{2}}\big(\ket{\hat{0} }\ket{1 } 
                           + \ket{\hat{1} }\ket{0 }\big)\,,\\
\ket{\varphi_4} &\defeq \frac{1}{\sqrt{2}}\big(\ket{\hat{0} }\ket{1 } 
                           - \ket{\hat{1} }\ket{0 }\big)\,,\\ \label{interfacestates}
\end{aligned}\end{equation}
but, according to Hawking's result, is in a one dimensional Hilbert space spanned by $\ket{\varphi_1}$. 

Consider the state of the system when $m$ qubits of radiation have already been emitted. In the computational basis the state of the system may then be written as
\be
\ket{BH+rad_m}=\sum C_{\hat q_1  \dots \hat q_{n+m} q_m  \dots q_1} \ket{\hat q_1  \dots \hat q_{n+m} } \otimes\ket{q_m  \dots q_1}\,. 
\ee
Mathur introduced small corrections to the Hawking's process in~\cite{Mathur:2009hf} by changing the evolution \bref{HawkingEvolution} to
\bea
&& \ket{BH+rad_m} \nn
 && \longrightarrow \sum C_{\hat q_1  \dots \hat q_{n+m} q_m  \dots q_1} ~~ \hat P_1\left( \ket{\hat q_1 \hat q_2 \dots \hat q_{n+m}} \right)~ \otimes ~\ket{\varphi_1^{n+m+1,m+1}} ~\otimes~\ket{q_m q_{m-1} \dots q_1} \nn
&& ~~+\sum C_{\hat q_1  \dots \hat q_{n+m} q_m  \dots q_1} ~~ \hat P_2 \left( \ket{\hat q_1 \hat q_2 \dots \hat q_{n+m}} \right)~ \otimes ~\ket{\varphi_2^{n+m+1,m+1}} ~\otimes~ \ket{q_m q_{m-1} \dots q_1} \nn
\eea
where note that $\hat P_1$ and $\hat P_2$ act only on the hatted qubits and the superscripts in $\ket{\varphi_i}$ denote the location of the newly created qubit pair for hatted and unhatted qubits respectively. He showed that as long as 
\be
\f{\langle \hat P_2^\dagger \hat P_2 \rangle}{\langle \hat P_1^\dagger \hat P_1 \rangle} = \epsilon \ll 1 \label{smallcorrections}\,,
\ee
where the expectation value in \bref{smallcorrections} is taken with respect to the state  $\ket{BH+rad_m}$,
the entropy of the radiation keeps increasing at every step, never turning around to make the final state pure as shown in Figure~\ref{evaporation}b. The condition \bref{smallcorrections} serves to keep the corrections to the leading process \bref{HawkingEvolution} small. Avery~\cite{Avery:2011nb} generalized this to include the other two directions in the space of Hilbert space of the two qubits~\bref{interfacestates} at the intersection completing Mathur's proof.

Simply stated, small corrections to the Hawking's process which respect effective field theory outside the black hole ($\hat P$s do not act on the unhatted qubits) lead to an ever increasing von-Neumann entropy of the radiation outside and thus have no hope of restoring unitarity.  Since Hawking's process arose because of the state at the horizon being in the Unruh vacuum, the above result can be restated as
\begin{quote}
Small corrections to the Unruh vacuum at the horizon maintain an ever increasing entropy of the radiation outside as shown in Figure~\ref{evaporation}b, thus precluding purity of the final state and therefore unitarity.
\end{quote}
Recently, Almheiri et. al.~\cite{Almheiri:2012rt} (AMPS) argued for the contrapositive of the above statement
\begin{quote}
Requiring the entropy of the radiation to go down after Page time, in accordance with Figure~\ref{evaporation}a, implies large deviations from the Unruh vacuum at the horizon {\em after Page time},
\end{quote}
to argue that an infalling observer would see drastic consequences on trying to fall through the horizon of an ``old" black hole. They  dubbed the horizon of an old black hole a firewall.\footnote{Relatedly, Ref.~\cite{Braunstein:2009my}, using a different argument, posits that an infalling observer hits an "energetic curtain" at the horizon of a black hole; however it is suggested that this phenomenon may only arise once the black hole is Planck-sized.} 

Note that ``old" black hole means a black hole entangled with some other system. While this system is the Hawking radiation in this case, it could be an arbitrary system. In the case of big black holes in AdS/CFT, the other system may be thought of as a system coupled to the CFT which sucks out the Hawking radiation leading to mixed state~\cite{Avery:2013exa,Almheiri:2013hfa}.

\subsection{Possibility of structure at horizon before Page time/ for pure states}

While one would expect that any structure at the horizon would be independent of how ``old" the black hole is, or more precisely whether the black hole is in a pure state or maximally mixed state, the nature of the argument in~\cite{Mathur:2009hf,Avery:2011nb,Almheiri:2013hfa} depends on just the purity of the final state of the radiation and thus the non-smoothness of the horizon can only be inferred after Page time.

In~\cite{Avery:2012tf} the author, along with Avery and Puhm,  argued that purity of final state is only one of the requirements of unitarity. Using other requirements for unitarity -- linearity, norm preservation and invertibility -- we argued that the horizon cannot be smooth even before Page time, i.e. for black holes in pure states. Our proof depends on the assumption that the initial black hole state is not special i.e. one can start in any of the $e^{S_{BH}}$ states. This is certainly not true for black holes formed from stellar collapse~\cite{tHooft:1993gx}. In such cases the Hilbert space of pre-collapse configurations are much smaller than  the Hilbert space of the black hole formed after collapse. Thus, such states are special in a sense. 

If black holes always formed in some special states, it is possible to speculate that the dynamics of black hole evaporation may be fine tuned to act on such special states in such as way so as to delay any significant deviation from the Unruh vacuum at the horizon till after Page time(see~\cite{Susskind:2012uw} for instance). In fact, one may even speculate non-local fine tuned dynamics to act on special states to keep the horizon predominantly in the Unruh vacuum.

However, in the next section we will see that there is no fundamental obstruction in creating black holes in generic states. This will put the proof of~\cite{Avery:2012tf} on a firm footing.

\section{The storage capacity of a black hole and its efficient utilization} \label{BHStorage}

Black hole entropy is huge compared to ordinary matter we encounter everyday. In particular it is given by the area of the black hole in Planck units
\be
S_{BH} = \f{A}{4G}\,.
\ee
Entropy is a measure of information and therefore of storage capacity. A natural question is whether all this storage space can be used. In other words, can a black hole be formed in any of the $e^{S_{BH}}$ states (or superpositions thereof) or is the actual storage capacity considerably limited (like an old 5 1/4" floppy disk with bad sectors)?

Black holes are usually thought of as formed from collapse of stars so the information content of a black hole formed from collapse is at best the amount of information in the pre-collapsed star. This amount was estimated in~\cite{tHooft:1993gx}. The most probable state of matter of energy $E$ in a ball of volume $V$ is a gas at temperature $T$ with the equation of state
\be
E = c_1 V T^4\,.
\ee
The entropy and therefore the logarithm of the number of possible pre-collapse configurations  of the star is
\be
S = c_2 V T^3\,.
\ee
The constants $c_1$, $c_2$ above and $c_3$ below are order one numbers. For the star to exist, it has to be bigger than its own Schwarzschild radius
\be
2 G E < (\f{3 V}{4 \pi})^\f{1}{3}\,.
\ee
Thus, we find the entropy of the star to be
\be
S_{matter} < c_3 (\f{A}{G})^\f{3}{4}\,.
\ee
Since the number of states is exponential in the entropy, it seems that while the black hole could in principle have existed in many different states, it actually exists in far fewer ones, at least for black holes formed by collapse of stars. When a state in a Hilbert space gets mapped to a bigger Hilbert space,  a corollary of the no-cloning theorem is that the map has to be of the form
\be
\ket{\psi} \to \ket{\psi} \otimes \ket{\Phi}
\ee
where $\ket{\Phi}$ is a fiducial state independent of $\ket{\psi}$~\cite{Avery:2011nb}. In  time this tensor product nature of the state will change as the state gets scrambled but it does not change the fact that the map is from a smaller Hilbert space to a bigger one and so these states are special in a sense. This has implications for the information paradox and the recent discussion about fuzzballs/firewalls.

In terms of qubit models,  the above discussion amounts to saying that the black hole is a $n$-qubit system and $n^{3/4}$ qubits contain some information of the state but the rest are in a fiducial state. For large $n$, the number of qubits carrying any actual information is a small fraction and one may imagine the black hole state to be more or less independent of the pre-collapse state. Then one may conjecture special dynamics for these small set of special states which allow a drama free experience at the horizon. So, it is very important to ask if black holes are limited to a exponentially suppressed number of total states available or they can be formed in any one of the $e^{S_{BH}}$ states.

\subsection{Schwarzschild black hole}

\begin{figure}[htbp]
\begin{center}
\includegraphics[scale=.2]{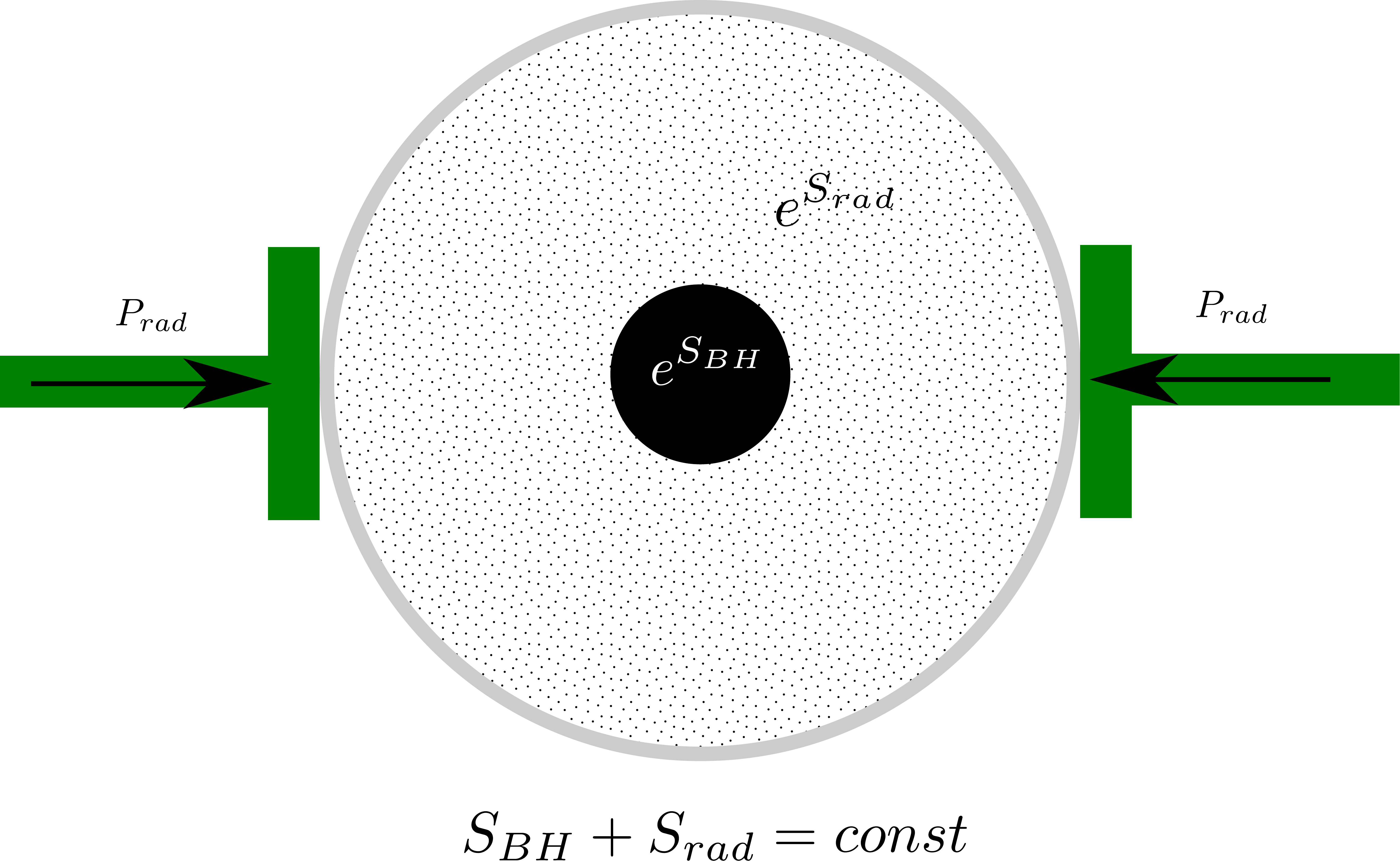}
\caption{A black hole when placed in a  perfectly reflecting container will come to equilibrium with its radiation under certain conditions. As long as such conditions are maintained, the box may be slowly expanded and contracted while keeping the total entropy of the system fixed.
}
\label{AdiabaticBHFormation}
\end{center}
\end{figure}

First we begin by looking at Schwarzschild black holes in $1+3$ dimensions and ask if we can arrange things such that
\be
S_{matter} \sim S_{BH} \label{equalEntropies}\,.
\ee 
In~\cite{PhysRevLett.49.1683}, Zurek showed that while the process of free streaming Hawking radiation produces more entropy in the radiation than is lost by the black hole\footnote{This process is not unique to black holes and happens free streaming radiation from any black body.}, the black hole may be evaporated in an adiabatic fashion so that the entropy gained by the radiation is equal to that lost by the black hole. The idea is to put a black hole in a perfectly reflecting box so that it is in equilibrium with its own radiation. The walls of the box can be connected to a piston so that it can be adiabatically expanded with no net entropy production. Once all, or at least a major fraction, of the energy has been converted into radiation one can tweak the state of the radiation to produce any one of the $e^{S_{BH}}$ states and adiabatically collapse the tweaked state again to form a black hole in any desired state. He  concluded that the states of the black hole correspond to pre-collapse configurations. This process is depicted in Figure~\ref{AdiabaticBHFormation}.

Let us revisit this problem. We take a system consisting of radiation of energy $E_{rad}$ and a black hole of energy $E_{BH}$ in a box of volume $V$. The entropies of the individual subsystems are
\begin{gather}
S_{rad} =\frac{4 \sqrt{\pi}}{3 (15)^\f{1}{4}} V^{1/4} E_{rad}^{3/4}\,, \qquad S_{BH} = 4 \pi G E_{BH}^2
\end{gather}
and the entropy of the total system is the sum of the entropies of the subsystems, $S= S_{rad} + S_{BH}$. The most stable configuration would be the one which maximized the entropy subject to the constraint $E= E_{rad} + E_{BH}$ being constant. These amount to~\cite{PhysRevD.13.191},
\begin{align}
\f{\partial S_{rad}}{\partial E_{rad}}  - \f{\partial S_{BH}}{\partial E_{BH}} =0 &~~~ \Longleftrightarrow ~~~T \defeq \left( \f{15 E_{rad}}{\pi^2 V} \right)^{1/4} = \f{1}{8 \pi G E_{BH}}\,,  \label{Temperature}\\
\f{\partial^2 S_{rad}}{\partial E_{rad}^2}  + \f{\partial^2 S_{BH}}{\partial E_{BH}^2} <0 & ~~~ \Longleftrightarrow ~~~ E_{rad} < \f{1}{4} E_{BH}\,. \label{ConditionOnEnery}
\end{align}
We can rewrite the total entropy and the condition for stable equilibrium \bref{ConditionOnEnery} in terms of the common temperature \bref{Temperature}
\be
S=  \f{1}{16 \pi G T^2}( \f{64 \pi^3 GVT^5}{45} + 1)\,, \label{Entropy}
\ee
and
\be
\f{32 \pi^3 G V T^5}{15} <1 \label{ConditionOnTemperature}
\ee
respectively.

Now we immediately see a problem. In order for the gas to have more entropy than the black hole we need
\be
GVT^5 \gg 1
\ee
which is inconsistent with the stability condition. So, contrary to the claims of Zurek, we can not adiabatically evaporate a Schwarzschild black hole sufficiently to confirm that the entropy of the black hole is a measure of pre-collapse configurations.

It is easy to see what the source of the above problem is. The black hole has a negative specific heat and when we expand the box adiabatically, at some point the black hole will become small enough so that the total specific heat will become negative. Then the black hole will not remain in stable equilibrium with its radiation. While it seemed bound to happen at some point, it was not obvious a priori that this will happen before the radiation can have more entropy than the black hole. 

However, just because this procedure does not work for Schwarzschild black holes, does not mean it will not work for other black holes as we will see next.

\subsection{D1-D5 black string}

The D1-D5 system has been very useful in the study of black holes. We will study this system for its storage capacity. One of the key differences of the near extremal D1-D5-P black string compared to the Schwarzschild black hole is that it has positive specific heat. Thus we may bypass the problem we faced in the preceding section. 
We give a brief description of this system sufficient for our purposes.  For a more in-depth review of the system see~\cite{Chowdhury:2010ct} for example. 

We compactify type IIB on string theory on $S^1 \times T^4$ with the four torus being string scale and the radius of $S^1$ being $R$. We wrap $n_1$ D1 branes on $S^1$ and $n_5$ D5 branes on $S^1 \times T^4$. We focus on the limit $R \gg g l_s$ where D-branes are a lot heavier than momentum along $S^1$ and any extra energy goes into exciting equal amounts of momentum and anti-momentum along $S^1$. The metric and other fields of this system may be found in full detail in~\cite{Giusto:2004id}. For our purposes it is sufficient to consider the system when there is no rotation in the non-compact direction and no net momentum. This simplified metric may be found in~\cite{Avery:2013exa}.

The energy, entropy and temperature in terms of the quantized charges are
\be
E_{BS} = \f{2 n_p}{R}\,, \qquad S_{BS} =4 \pi \sqrt{n_p n_1 n_5}\,, \qquad T_{BS} = \f{1}{\pi R} \sqrt{\f{n_p}{n_1 n_5 }}\,.
\ee
The above expression for the entropy can be obtained from the gravity as well as field theory~\cite{Horowitz:1996fn}.  

The system will behave differently for high and low temperatures but as we will see we need only worry about high temperatures. At high temperatures ($T_{rad}R \gg 1$) the box is effectively $5+1$ dimensional and the energy and entropy of radiation is given by
\be
E_{rad} = c_4 V R T_{rad}^6\,, \qquad S_{rad} = c_5 V  R T_{rad}^5
\ee
where $c_4$ and $c_5$ are some order one constants. This makes the total entropy 
\be
S= c_2 V R \left( \f{E_{rad}}{c_1 VR} \right)^{5/6} + 2 \pi \sqrt{2 R n_1 n_5 E_{BS}}\,.
\ee
The condition for equilibrium is that the subsystems have the same temperature
\be
T \defeq \left( \f{E_{rad}}{c_1 V R} \right)^{1/6} =  \f{1}{\pi R} \sqrt{ \f{E_{BS} R}{2 n_1 n_5}}\,.
\ee
This equilibrium is always stable as 
\bea
\f{\partial^2 S_{rad}}{\partial E_{rad}^2}  + \f{\partial^2 S_{BS}}{\partial E_{BS}^2}  &=& - \left( \frac{5  c_2}{36  c_1^2 R V T^7}+\frac{1}{4 \pi^2 n_1 n_5   R  T^3} \right) <0
\eea
so we do not expect problems like we had with the Schwarzschild black hole in adiabatically evaporating the black string.

As we increase the volume of the surrounding box adiabatically, the temperature will decrease. There is evidence from the weakly coupled D1-D5 system at the orbifold point that around $T R \sim 1$ there will be a phase transition coming from the long string sector converting into shorter strings~\cite{Mathur:2005ai}. Further, when $TR \sim 1$ the surrounding box becomes $4+1$ dimensional so the radiation system also undergoes a phase transition. Fortunately we will not have to worry about these subtleties since by that time the radiation will have much more entropy than the black string as we will see.

In order to see this we first write the entropy in terms of the common temperature
\be
S  =4 \pi^2  RT( \f{c_2}{4 \pi^2} \f{V}{R^4}  (RT)^4 +  n_1 n_5 )\,.
\ee 
We want to start with a high temperature configuration such that most of the entropy is in the black string. The entropy is then
\be
S \approx  RT_i n_1 n_5\,. \label{InitialEntropy}
\ee
We want to adiabatically expand the box till the temperature is order one $T_f R \sim 1$. The entropy is the same as before but in terms of the final volume is
\be
S \approx \f{V_f}{R^4} + n_1 n_5 \label{FinalEntropy}\,.
\ee
Equating \bref{InitialEntropy} and \bref{FinalEntropy} and using $R T_i \gg 1$ we find
\be
\f{V_f}{R^4} \approx RT_i n_1 n_5 \gg n_1 n_5
\ee 
so most of the entropy is in the radiation by the time $TR \sim 1$. One can now change the state of the radiation and adiabatically pump it back into the D1-D5 system producing it in any of the $e^{S_{BS}}$ states. 

We expect this to work for all black holes with positive specific heat. Specifically, it should work for all big black holes in AdS. After all, in that case all we are saying is that the dual CFT can be put in any one of the $e^{S_{BH}}$ states accounting for its microcanonical entropy and there does not seem to be any fundamental obstruction to this.

\section{The cost of cool horizons} \label{StructurefulHorizons}

In the previous section we established that it is indeed possible to make black holes in any one of $e^{S_{BH}}$ states. This process will be slow but that should not in any fundamental way affect the properties of the black hole. In this section we will be talking about such black holes in arbitrary states. We will review an earlier result~\cite{Avery:2012tf} which showed that such black holes cannot have Unruh vacuum at the horizon even before Page time. We will then end by demonstrating how some of the arguments against fuzzballs/firewalls do not refute the results of Ref.~\cite{Avery:2012tf}.

\subsection{Unitarity requires order one corrections at each step of emission} \label{StructurefulHorizonsOldResult}

Since the argument about small corrections made by Mathur and Avery was based on purity (or lack thereof) of the final state of the radiation from a black hole, the contrapositive of the statement could only be made after Page time. This has resulted in a lot of discussion about ``young" vs. ``old" black holes. Expecting this to be a red herring, the author along with Avery and Puhm revisited the problem~\cite{Avery:2012tf}. Instead of focussing merely on purity of final radiation, which is necessary but not sufficient for unitarity, we analyzed other requirements -- linearity, norm-preservation and invertibility. Either by appealing to finiteness of the black hole Hilbert space (which appears violated in the nice slice description but is necessary in a unitary theory) or by appealing to energy conservation one can argue that the process of radiation from an $n$-qubit system terminates after  $n$ steps (see footnote~\ref{numberOfEmissions}). Thus, all the information of $n$ qubits have to be mapped onto the outside Hilbert space in $n$ steps. This allows very little wiggle room. At each step one qubit worth of information has to be removed from the inside Hilbert space and mapped to the outside Hilbert space.

The proof of this rather intuitive idea is technical and we will not repeat it here. The interested reader can find it in Ref.~\cite{Avery:2012tf}. Fortunately, the result can be stated quite simply:
\begin{quote}
The typical state of the interface of two interacting systems, at least one of which is local (the outside), can not preferentially be in a subspace of the the Hilbert space spanned by \bref{interfacestates}. Instead, the typical state necessarily explores the full four dimensional Hilbert space.
\end{quote}

Our result is stronger than the results of~\cite{Mathur:2009hf,Avery:2011nb} and proves that there has to be an order one correction to the Unruh vacuum at the horizon even for young black holes. At this point one may choose to say that a firewall has to be there from the beginning but since the main point of the fuzzball proposal is that there is structure at the horizon precluding universal infall, our result strongly supports the fuzzball conjecture.

In fact, it is not obvious how one should see firewalls as distinct from fuzzballs (see also~\cite{Almheiri:2013hfa}). The distinction seems to originate from confusing the fuzzball proposal with fuzzball complementarity in the original AMPS paper. 
The fuzzball proposal argues for structure at the horizon which would change the evolution of typical ($E \sim T_{BH}$) quanta by order one and is agnostic about the infall of non-typical quanta. This proposal is primarily a constructive one with many string theoretic microstates of gravity having been  systematically found over the last decade(see~\cite{Mathur:2005zp,Bena:2004de,Skenderis:2008qn,Balasubramanian:2008da,Chowdhury:2010ct} for reviews). The latter, on the other hands, is a relatively recent proposal which states that non-typical infalling quanta may still experience a universal behavior described by free infall. In this article we will not say much about fuzzball complementarity as our goal is to argue that structure at the horizon -- fuzzballs -- are present from very early on, possibly being delayed till scrambling time. For a recent discussion on fuzzball complementarity, especially in relation to AMPS, see~\cite{Mathur:2013gua}. In this paper we shall refer to order one deviation from the Unruh vacuum as fuzzballs mainly and occasionally as firewalls when describing responses to AMPS.

\subsection{Why some ``resolutions" of our result/fuzzballs/firewalls do not work} \label{StructurefulHorizonsProblemWithOtherModels}

The results of~\cite{Avery:2012tf} are in direct contradiction with some papers claiming to resolve some or all issues with fuzzballs/firewalls so we will address three such models out of the burgeoning literature.  

There are two results in~\cite{Hawking:1974sw} which show violation of unitarity during black hole  evaporation:
\begin{itemize}
\item A pure initial state evolves to a mixed state,
\item The map from the initial state to the final state is many to one and thus non-invertible.
\end{itemize}
Thus, even if one proposes a model, with or without a mechanism to bypass Hawking's pair production, that ends up giving a final state which is pure,  it still needs to be checked for invertibility of the map from the initial to the final state. We will see that the proposed resolutions that we discuss, end up making the horizon smooth at the cost of unitarity.

The proofs of Mathur~\cite{Mathur:2009hf}, Avery~\cite{Avery:2011nb} and AMPS~\cite{Almheiri:2012rt} only address the first issue above and boil down to using strong subadditivity on three systems. Following the notation of these three papers they are $A$ -- the early radiation, $B$ -- the outgoing hawking quantum and $C$ -- the infalling hawking quantum. We will not repeat the proofs here but the punch line of the argument forwarded by AMPS is that purity of the final state implies the horizon cannot be pure after Page time
\be
S_{AB}< S_A ~~\Rightarrow ~~S_{BC} \ne 0  \qquad  \forall ~ t>t_{\text{Page}}\,.
\ee
In particular, this implies the horizon cannot be in the Unruh vacuum which is pure
\be
S_{BC} \ne 0  ~~~ \Rightarrow ~~~ \ket{BC} \ne \ket{\varphi_1}  \label{SBCZero1}\,.
\ee
where recall that in the qubit models $\ket{\varphi_1}$ is the Unruh vacuum~\bref{QubitUnruhVacuum}. A lot of responses to AMPS have focussed on finding a way to make the $BC$ system pure in such a way that $B$ is maximally entangled with $C$. The problem with this approach is that the Unruh vacuum is not the only state with this property.\footnote{We thank Steve Avery for helpful discussions on this point.} In particular this property is shared by all the four Bell states \bref{interfacestates}  and only one of them, $\ket{\varphi_1}$,  plays the  role of the Unruh vacuum \bref{QubitUnruhVacuum}. The others are an order one deviation and cannot allow free infall. More explicitly,
\be
S_{BC} = 0  ~~~ {\not \Rightarrow} ~~~ \ket{BC} = \ket{\varphi_1} \label{SBCZero2}\,.
\ee
\\
In fact, the most general two qubit state with $S_{BC}=0$ and with $B$ and $C$ maximally entangled is of the form
\be
\ket{\eta_{\theta,\phi,\chi}}=\f{1}{\sqrt{2}} \Bigg[  \left( \cos \Big(\f{\theta}{2}\Big) ~\ket{\hat 0} +e^{i \phi} \sin \Big(\f{\theta}{2}\Big)~ \ket{\hat 1} \right)) \ket{0} + e^{i \chi} \left( \sin \Big(\f{\theta}{2}\Big) ~\ket{\hat 0} - e^{i \phi} \cos \Big(\f{\theta}{2}\Big)~ \ket{\hat 1} \right) \ket{1} \Bigg]   \label{GeneralBC}
\ee
where $0 \le \theta \le \pi, 0 \le \phi,\chi < 2 \pi$. We will refer to these as generalized Bell pairs.
The fact that $S_{BC}=0$ does not imply that $\ket{BC}=\ket{\varphi_1}$ and instead only implies that $\ket{BC}$ is one of the generalized Bell pairs \bref{GeneralBC} will be the central theme in the problems with the models we discuss below.\footnote{Note it is not correct to {\em define} the hatted qubits so that $\ket{BC} = \f{1}{\sqrt{2}} ( \ket{\hat 0 0} + \ket{\hat 1 1})$ qubit pair by qubit pair. This is because the Unruh vacuum is defined by the physics of infall which fixes the definition once and for all.}

\subsubsection*{Model 1 - Distillable entanglement model}

In~\cite{Susskind:2012uw} Susskind explains that if we take a Hilbert space of size $2^n$ and divide it into system $A$ of size $2^m$ and $B$ of size $2^{n-m}$, any typical state $\ket{\psi}$ on $A \cup B$ may be written as
\be
\ket{\psi} = U_A \otimes U_B \left[ \f{ \ket{0_A} \ket{0_B} +\ket{1_A} \ket{1_B}}{\sqrt{2}} \right]^{Min(m,n-m)}
\ee
for $n,m,n-m \gg 1$, where $U_A, ~U_B$ are unitary operations on $A$ and $B$ respectively. This follows from the result of Page~\cite{Page:1993df,Sen:1996ph}. What has been done here is that the entanglement between system $A$ and system $B$ has been first expressed in terms of Bell pairs \bref{interfacestates} (or more precisely in terms generalized Bell pairs  \bref{GeneralBC}) and then these have been rotated into the first Bell pair $\ket{\varphi_1}$. A count of how much two systems are entangled is the distillable entanglement. However, that entanglement generically is not in the form of $\ket{\varphi_1}$ and to convert all other generalized Bell pairs to $\ket{\varphi_1}$ requires a {\em state dependent} transformation.

One can then posit that the dynamics of the black hole is such that it takes a typical state and acts on it with $(U_A \otimes U_B)^{-1}$ for equal sized systems so that $m=n/2$. 
\be
\ket{BH} = (U_A \otimes U_B)^{-1} \ket{\psi} = \left[ \f{ \ket{0_A} \ket{0_B} +\ket{1_A} \ket{1_B}}{\sqrt{2}} \right]^{n/2}\,. \label{SusskindMap}
\ee
Then before Page time quanta from system $A$ are emitted in such a way that the interface is always in the state
\be
\f{ \ket{0_A} \ket{0_B} +\ket{1_A} \ket{1_B}}{\sqrt{2}}\,.
\ee
This will allow free infall. However, after Page time this process has to stop as now quanta from system $B$  will be emitted and there can be no free fall. Thus, in~\cite{Susskind:2012uw} Susskind argued that firewalls can be avoided before Page time but not after.

One problem with this kind of argument, as mentioned above, is that $(U_A \otimes U_B)^{-1}$ is  state dependent
\be
(U_A \otimes U_B)^{-1} = (U^{\ket{\psi}}_A \otimes U^{\ket{\psi}}_B)^{-1}\,.
\ee
A worse problem is the fact that it is a non-invertible map. Recall that the information problem stemmed from a many to one non-invertible map in the first place.

If this is the model of black hole evaporation then the question is how would Bob standing outside the black hole collecting all  the radiation coming from the black hole construct the original state $\ket{\psi}$. A system which can perform the map \bref{SusskindMap} would have to be at least as big as the black hole and according to the no-hiding theorem~\cite{Braunstein:2006sj}, would have absorbed all the information. After the black hole evaporates, Bob has access to the radiation but it carries no information. Bob would need access to the system which bleached out the black hole to produce the map \bref{SusskindMap} but there is no such system left.

So in summary the problem with this model is that it restores purity but still looses unitarity. In fact a model equivalent to this was explicitly constructed in~\cite{Avery:2012tf} to show how it does {\em not} restore unitarity even though the final radiation is pure.

\subsubsection*{Model 2 - ``A=C" models}

Another argument voiced by many, based in no small extent on the original idea of black hole complementarity, is that the inside Hawking quantum $C$ should not be thought of as distinct from the early radiation quanta $A$. The claim is that this invalidates a key assumption in the proofs of~\cite{Mathur:2009hf,Avery:2011nb,Almheiri:2012rt}.

Papadodimas and Raju tried to make this argument concrete by proposing a model based on this idea in~\cite{Papadodimas:2012aq}.  They envision the following dynamics for infall. When Alice encounters a qubit $B$, it is maximally entangled with the rest of the qubits when the full state is a typical state in the Haar measure sense. However, the entanglement entropy is $\log 2$ so $B$ is entangled with {\em one} non-locally spread out qubit $C$ such that $S_{BC}=0$.

Papadodimas and Raju posit infall dynamics such that it is this non-local $C$ which Alice encounters right after $B$. They claim that this will make Alice's experience fuzzball/firewall free both before and after Page time. The problem with this is of course that $S_{BC}=0$ is not enough to say the state is the Unruh vacuum as explained in the beginning of this section and mentioned explicitly in \bref{SBCZero1} and \bref{SBCZero2}.

To say that the observer always encounters $BC$ such that it is in the state $\ket{\varphi_1}$ implies that the {\em non-local} dynamics involves the bleaching operation
\be
\ket{\eta_{\theta,\phi,\chi}} \to \ket{\varphi_1}\,. \label{ManyToOne}
\ee 
which converts all generalized Bell pair states \bref{GeneralBC} to the Unruh vacuum $\ket{\varphi_1}$.
Thus we again have a non-invertible map like in the previous model and the cost we pay for smooth horizons is information loss.

\subsubsection*{Model 3 - ``ER=EPR" model}

A variant of the firewall argument may be made by replacing the Hawking radiation by an arbitrary heat bath which is entangled with the CFT~\cite{Avery:2013exa,Almheiri:2013hfa,VanRaamsdonk:2013sza}.

A similar idea was recently discussed by Maldacena and Susskind in~\cite{Maldacena:2013xja} but with the opposite conclusion. To begin, note that Maldacena~\cite{Maldacena:2001kr} has proposed that  a system of two CFTs which are entangled in a particular way
\be
\ket{EBH} = \f{1}{\sqrt{Z}} \sum e^{-\beta E/2} \ket{E}_L \otimes \ket{E}_R   \label{MaldacenaState}
\ee
is dual to the eternal AdS black hole. 

Maldacena and Susskind argue that it is possible for Alice to collect all the radiation of an old black hole and collapse it to form another black hole. While these two black holes will be maximally entangled they might not be in the state~\bref{MaldacenaState}. In this case the horizons will not be ``cool". This is the analogue of our \bref{SBCZero1} and \bref{SBCZero2}. However, Maldacena and Susskind claim that if Alice has an extremely powerful quantum computer she can have it do a quantum operation to put the two black holes in the state~\bref{MaldacenaState} leading to ``cool" horizons.

\begin{figure}[htbp]
\begin{center}
\subfigure[]{
\includegraphics[scale=.15]{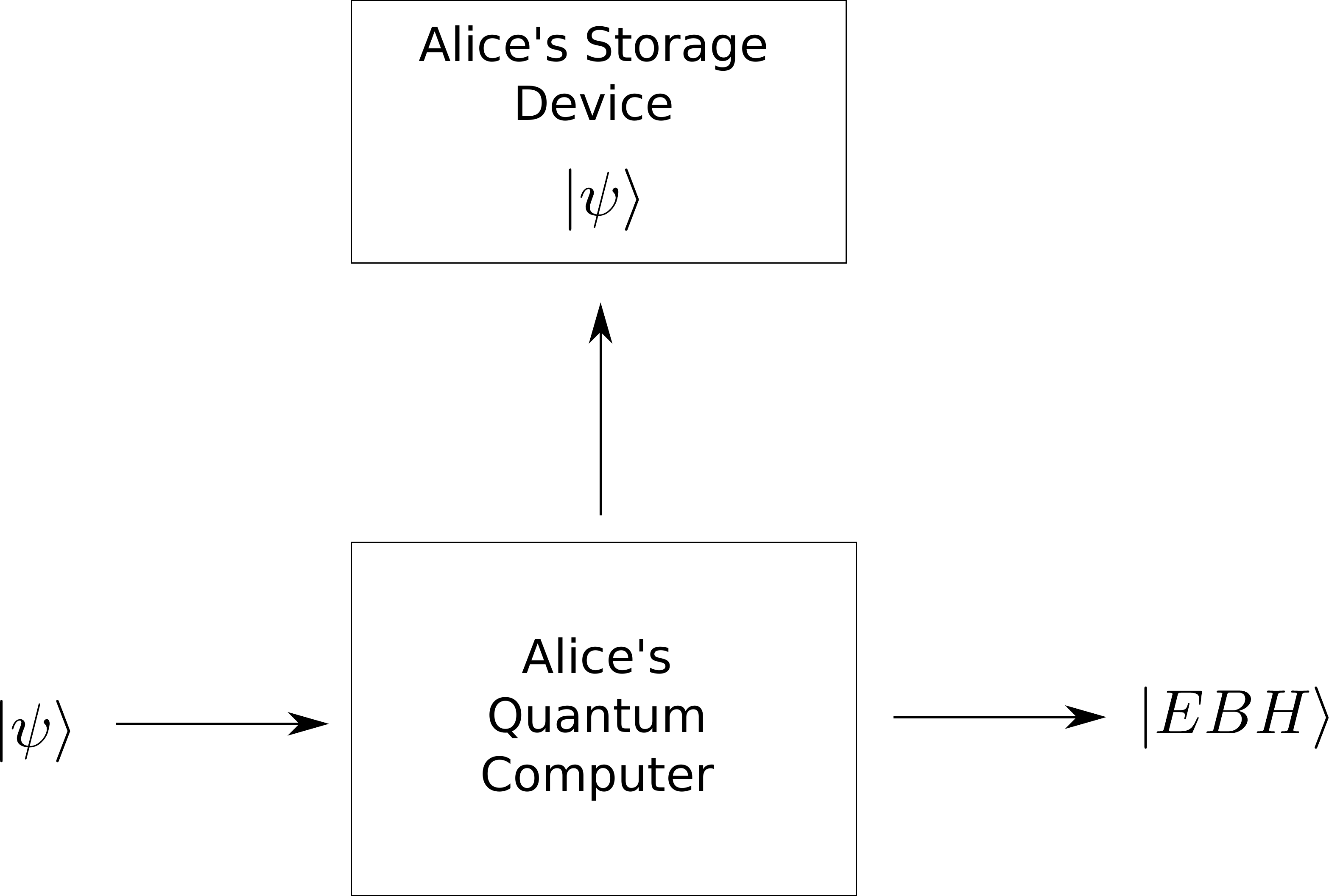}} \hspace{1in}
\subfigure[]{ 
\includegraphics[scale=.15]{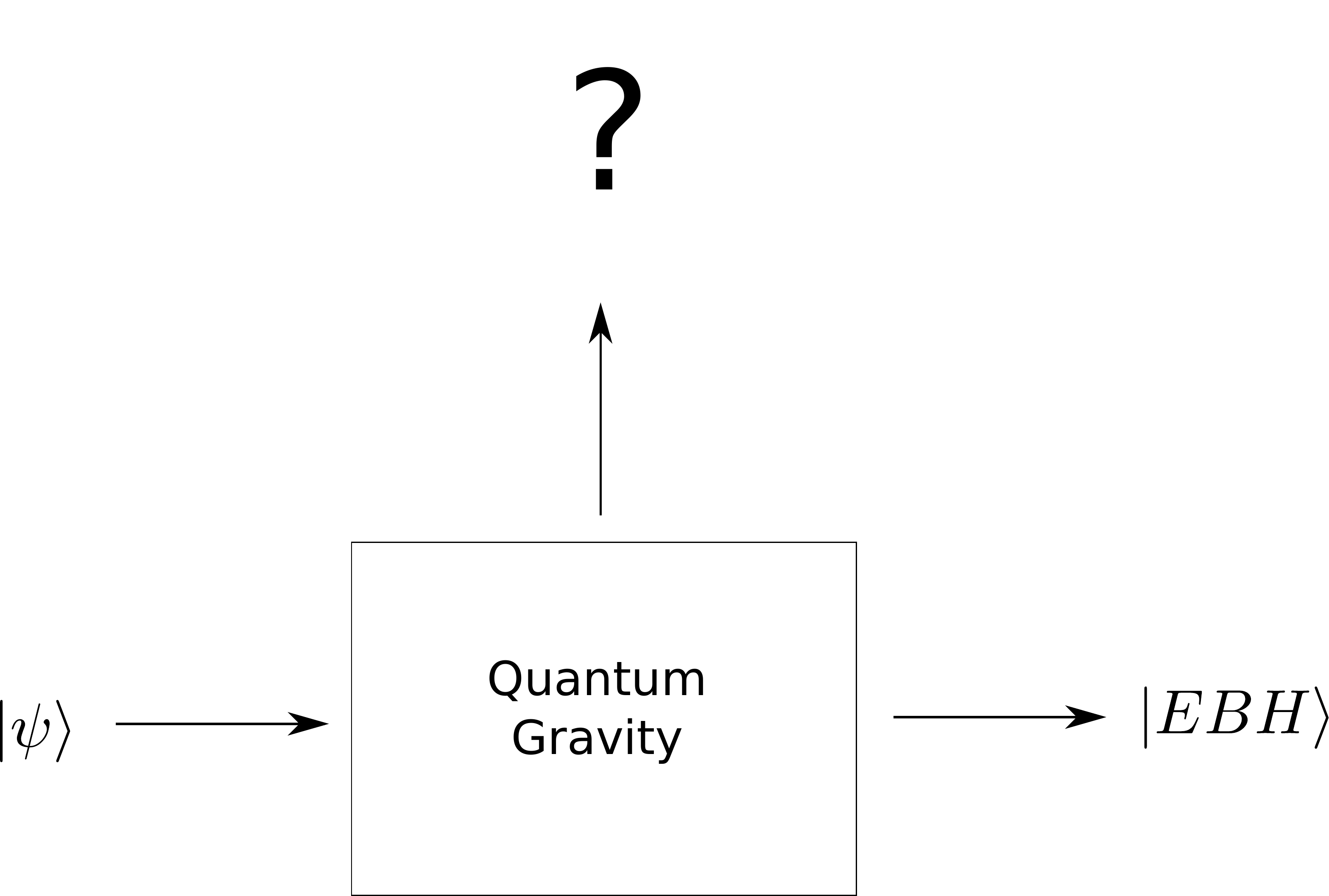} } \\
\caption{(a) Alice can bleach black hole and radiation system to the eternal black hole state using a powerful quantum computer. However, if this has to happen for generic states then the quantum computer has to store the state of the original system in some storage device the size of a black hole. (b) In the context of the information paradox we are more interested in what quantum gravity does than what Alice can do. If quantum gravity is to repeat Alice's task than it needs a separate storage device the size of the black hole which is bleached. In the absence of such a storage device we are back where we started -- information loss.
}
\label{Computers}
\end{center}
\end{figure}

While we agree with the above, we would like to add something to this. Of course, if Alice has access to the original black hole and radiation,  she can have her extremely powerful quantum computer do a quantum bleaching operation on the total system to put it in the state~\bref{MaldacenaState}. However, by unitarity such an operation would  require the quantum computer to {\em extract all the information of the original state onto a storage device}~\cite{Braunstein:2006sj}.  If this has to work for generic states the size of the storage device has to be at least as big as $e^{S_{BH}}$. This is shown in Figure~\ref{Computers}a.

So we see that in order to convert the original black hole and radiation into the eternal black hole state, Alice has to have access to another storage device as big as the black hole to save the information of the original state into. However, for the information loss problem we are not interested in what Alice can do but what quantum gravity does. If instead of Alice, {\em quantum gravity} has to do this quantum bleaching operation on  generic states of the black hole, then  what is this extra storage space? This is shown in Figure~\ref{Computers}b. If such a storage space does not exist (or is not accessible to asymptotic observer Bob), as seems to be the case then we are back with information loss.\footnote{If one were tempted to postulate such a storage in quantum gravity then one can simply postulate it for the original model of black hole evaporation without a need for models which give a pure final state. In the presence of an adequately large storage a pure-to-mixed map simply implies perfect thermalization with respect to the storage. Of course, one then has to explain what degrees of freedom are responsible for this extra storage.} 

\subsection{Relation to earlier work refuting $A=C$ models}

It was observed by  Bousso earlier in~\cite{Bousso:2012as} that $A=C$ models for generic maximally entangled states require a many to one map.\footnote{We thank Raphael Bousso for pointing this out to us.}  This was later also mentioned in~\cite{Bousso:1532382} where he noted that there seems to be problems with linearity because of this.


While our work in this section has some overlap with the technical findings of Bousso's work, but our paper goes beyond this. Though he claims that for many to one maps there seem to be problems with linearity, this misses a very important physical implication of non-invertibility.

As we discussed in detail above, such non-invertible maps are quantum operations which bleach a given Hilbert space. Such bleaching operations, if they are to preserve unitarity, need to transfer the information of the original state on the Hilbert space to some other Hilbert space (see Figure~\ref{Computers}). While one can certainly imagine such operations being performed by Alice with access to a huge storage, in quantum gravity such a storage is missing and any model which proposes such maps are basically advocating information loss. Of course if one advocates that information is lost, then there is no reason to even assume the final state is pure and Hawking's original model is fine.

\section{Conclusion and discussion} \label{Discussion}

Information loss paradox cannot be solved by simply proposing a model that makes the final state of the radiation pure. The map from the initial to the final state has to be invertible. We had demonstrated in~\cite{Avery:2012tf} that the latter requirement implies order one deviations from the Unruh vacuum during the entire evaporation of a black hole. A  loophole in our proof could be found if black holes only forms in special states. If such were the case, one could postulate special dynamics for special states which could keep the horizon predominantly in the Unruh vacuum. However, in this paper we have established that it is indeed possible to make black holes in any one of the $e^{S_{BH}}$ states.

 In light of this, models of~\cite{Susskind:2012uw,Maldacena:2013xja,Papadodimas:2012aq} may be interpreted in two ways: (i) they only allow a fuzz/fire free horizon for special states, (ii) they work on all states to allow fuzz/fire free horizons but at the cost of unitarity. Neither of these is a pleasing option. The problems with the first option are justifying the fine tuned dynamics for black holes formed from astrophysical collapse and treating black holes formed by fast and adiabatic collapse differently. Occam's razor would suggest that there is fuzzy structure at the horizon throughout a black hole's lifetime. The problem with option two is of course that we are back where we started -- information loss.
   
We are of the opinion that the results of this paper (see also~\cite{VanRaamsdonk:2013sza}) strongly support the idea that the horizon of black holes are not smooth and instead the geometries end in fuzzy-stringy states outside the horizon i.e. fuzzballs. See~\cite{Mathur:2005zp,Bena:2004de,Skenderis:2008qn,Balasubramanian:2008da,Chowdhury:2010ct} for some reviews of this proposal.

We end with a few general comments on why the black hole picture seems to work in so many respects despite being so problematic as far as unitarity is concerned.

\subsection{Analytic continuation and thermo-field doubling}

The technique of thermo-field doubling (TFD)~\cite{doi:10.1142/S0217979296000817} is used as a calculational tool in thermal field theory. If one has a density matrix in a thermal state of system ``Right"
\be
\rho_R = \f{1}{Z} \sum e^{-\beta E} \ket{E}_R \bra {E}_R\,, \label{densityMatrix}
\ee
one can {\em formally} purify this mixed state by doubling the Hilbert space by adding system ``Left" which is identical to the system ``Right" and writing the purified state as
\be
\ket{TFD} \defeq \f{1}{\sqrt Z} \sum e^{-\beta E/2 } \ket{E}_L \otimes \ket{E}_R\,. \label{PurifiedState}
\ee
Note that one recovers \bref{densityMatrix} from \bref{PurifiedState} by tracing over the system ``Left".
This procedure has an intimate relationship with analytic continuation. The eternal Schwarzschild black hole is obtained by maximally analytically continuing the Schwarzschild solution on the right, beyond the horizon. Israel observed that the left side of such a black hole is a TFD of the right side~\cite{Israel:1976ur}. In fact this result also applies to the two Rindler wedges obtained by Rindler decomposition of Minkowski space. Using ideas from AdS/CFT, Maldacena proposed that the eternal AdS black hole is dual to a system of two CFTs in a TFD state~\cite{Maldacena:2001kr}.

However, \bref{PurifiedState} is not the only purification of \bref{densityMatrix}. For simplicity we discuss these ideas using qubits. Let us look at the density matrix
\be
\rho_{unhatted} = \h ( \ket{0}\bra{0} + \ket{1}\bra{1})\,.  \label{densityMatrixQubit}
\ee
The TFD method would tell us that the purified state is
\be
\ket{TFD}  = \f{1}{\sqrt 2} ( \ket{\hat 0} \ket{0} + \ket{\hat 1} \ket{1})\,. \label{PurifiedStateQubit}
\ee
We certainly recover \bref{densityMatrixQubit} by tracing over the hatted Hilbert space in \bref{PurifiedStateQubit}. However, we also recover the \bref{densityMatrixQubit} by tracing over the hatted Hilbert space in any of the following
\begin{align}
\f{1}{\sqrt 2} ( \ket{\hat 0} \ket{0} - \ket{\hat 1} \ket{1})\,, \qquad  \f{1}{\sqrt 2} ( \ket{\hat 1} \ket{0} + \ket{\hat 0} \ket{1})\,, \qquad\f{1}{\sqrt 2} ( \ket{\hat 1} \ket{0} - \ket{\hat 0} \ket{1})\,,
\end{align}
among a continuum of other possibilities~\bref{GeneralBC}. In fact, we can also purify \bref{densityMatrixQubit} as
\be
\ket{\text{\cancel{TFD} purification}}  = \left( \sum c_{12\dots n} \ket{\hat q_1 \hat q_2 \dots \hat q_n} \right) \otimes \left[ \f{1}{\sqrt 2} ( \ket{\hat 0} \ket{0} + \ket{\hat 1} \ket{1}) \right] \label{PurifiedStateQubit}
\ee
with $n$ arbitrary. How do we understand this ambiguity in ``purification"? The point is that any mixed state will be mixed with something (its environment/ancillia/heat bath) and without access to the rest of the system, it is simply not possible to predict what the full state is. The TFD technique {\em assumes} that the system is mirrored in the heat bath.\footnote{This was also discussed in~\cite{Papadodimas:2012aq} but the claim was that the system {\em is} mirrored in the heat bath. This crucial point is what leads to the incorrect claim that smooth horizons are compatible with unitarity as discussed in Section~\ref{StructurefulHorizonsProblemWithOtherModels}.}

\begin{figure}[htbp]
\begin{center}
\subfigure[]{
\includegraphics[scale=1]{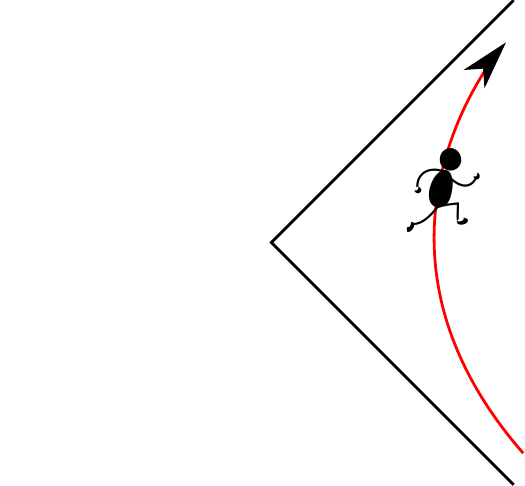}} \hspace{1in}
\subfigure[]{ 
\includegraphics[scale=1]{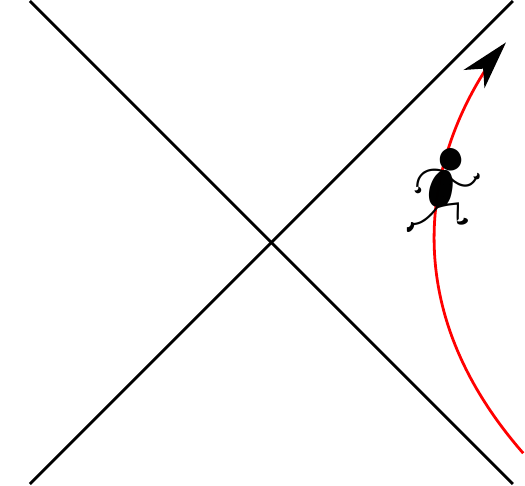} } \\
\subfigure[]{
\includegraphics[scale=1]{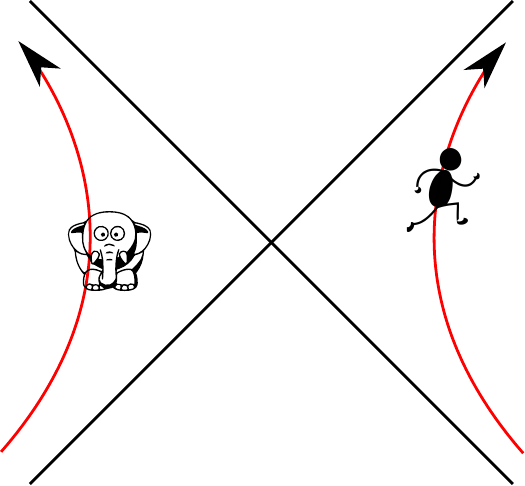}} \hspace{1in}
\subfigure[]{ 
\includegraphics[scale=1]{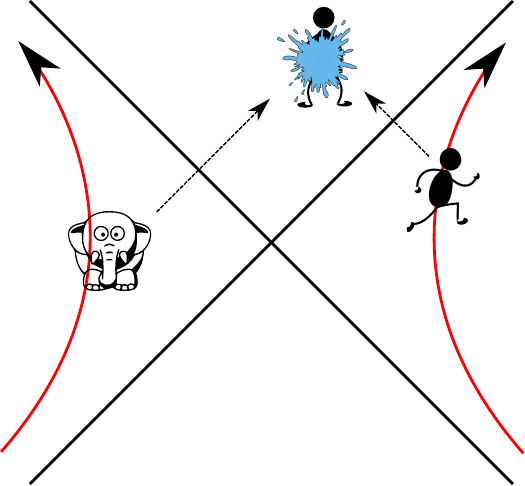} }
\caption{(a) Bob who is accelerating, experiences Rindler space which is described by a thermal density matrix. Such a spacetime has a horizon.  (b) Bob may use analytic continuation to {\em assume} that the full spacetime is the Minkowski vacuum.  Alternately the thermo-field double of his density matrix corresponds to the Minkowski vacuum. (c) However, the actual state may have a Rindler (i.e. accelerating) elephant in the left wedge. (d) While Bob will not pay for his mistake in using analytic-continuation/TFD, Alice who listens to Bob and jumps through Bob's horizon may end up being hit by a firewall of water thrown by the elephant.
}
\label{AnalyticContinuation}
\end{center}
\end{figure}

For instance, let us assume an accelerating Bob sees the thermal state~\bref{densityMatrix}. While a purification of the kind \bref{PurifiedState} corresponds to analytic continuation of Bob's wedge giving Minkowski vacuum, another possible purification contains  a Rindler (accelerating) elephant in the left wedge. These scenarios are shown in Figure~\ref{AnalyticContinuation}. Being accelerated, Bob will see a horizon in addition to the thermal state \bref{densityMatrix} as shown in Figure~\ref{AnalyticContinuation}a. If he uses analytic continuation or TFD purification he would expect the full state to be the Minkowski vacuum state \bref{PurifiedState} as shown in Figure~\ref{AnalyticContinuation}b. However, suppose the actual state is in fact the one with a Rindler elephant in the left wedge, as shown in Figure~\ref{AnalyticContinuation}c. Bob would not know that and will tell Alice that she may safely fall through the horizon. When Alice actually does so, she may get hit by a {\em firewall of water} splashed by the Rindler elephant in Figure~\ref{AnalyticContinuation}d.


The moral of the story is that by just having access to the thermal density matrix~\bref{densityMatrix} Bob is not justified in saying the full state is the thermo-field double state. Alternately, Bob is not justified in doing an analytic continuation. 

\subsection{Why does black hole counting work {\em or} what about the successes of semi-classical gravity?}

The discussion above tells us that if unitarity is to be preserved, quantum gravity effects become important at  horizon scale. Unitarity dictates that there is no hairless horizon and no interior of a black hole. Of course without these two, the phrase black hole does not make sense and the above discussion tells us that the true microstate of quantum gravity are fuzzballs which end in a quantum-fuzzy-stringy mess before the horizon. 

However, horizon scale is parametrically larger than Planck scale where one would naively have expected quantum gravity to become important.  One question which has been asked in the context of fuzzballs is --  if this is the case how does one explain success of semi-classical gravity like black hole state counting~\cite{Sen}? Here we will attempt to answer this question.

There are two ways to measure the entropy of a system (among others) -- in the microcanonical ensemble and in the canonical ensemble. In the microcanonical ensemble one actually counts the number of microstates consistent with the given macroscopic charges. In the canonical ensemble on the other hand, the entropy is given by
\be
S= \beta (E-F) = (\beta \partial_\beta -1)(\beta F)\,.
\ee
In the canonical ensemble the density matrix is thermal $\rho=e^{-\beta H}$ from which we get
\be
S=(\beta \partial_{\delta \beta} -1) [- \log Tr \rho^{1+ \delta \beta/\beta}] \Big |_{\delta \beta =0} = - Tr [\hat  \rho \log \hat \rho]
\ee
where $\hat \rho = \f{\rho}{Tr \rho}$ is the normalized density matrix. So the entropy in the canonical ensemble is the von-Neuman entropy. Thus, we see that the canonical ensemble entropy measures how much {\em the system is entangled with the heat bath}. One may then think that the states counted in the microcanonical ensemble end up getting entangled with the heat bath in the canonical ensemble in the thermodynamic limit.

Now, suppose we want to ask how many fuzzballs exists for a certain energy. For specificity we consider asymptotically AdS fuzzballs which are supposed to be dual to the CFT microstates. We can either count them one by one\footnote{This Herculian task is in fact being attempted by Bena, Warner and their collaborators~\cite{Bena:2007kg,Bena:2008nh,Bena:2010gg,Bena:2011uw,Bena:2011dd,Niehoff:2013kia,Shigemori:2013lta}.} or we can couple the fuzzballs to a heat bath and count entanglement. It turns out that for gravitational thermal density matrices there is an easy way to count entanglement.

From the discussion above we learnt that a thermal state may be purified in many different ways. The TFD purification is the simplest in many ways as it models the heat bath as another copy of the system. If we take the TFD purification of a thermal ensemble of fuzzballs, Van Raamsdonk argued in Ref.~\cite{VanRaamsdonk:2010pw,Czech:2012be,VanRaamsdonk:2013sza} that we get the eternal AdS black hole. For mysterious reasons, the Bekenstein-Hawking entropy of the eternal AdS black hole counts the entanglement entropy of one CFT with its thermofield double~\cite{Maldacena:2001kr}. Thus, the Bekenstein-Hawking entropy is a measure of entanglement of a thermal ensemble of fuzzballs with its heat bath.

To summarize, the Bekenstein Hawking entropy counts the entanglement entropy of a thermal ensemble of fuzzballs with its heat bath. If we had chosen to purify the thermal ensemble to a state other than the TFD we would not have had a geometric interpretation and would not have been able to obtain the answer so elegantly. However, that operation would not have effected the entanglement entropy. Also note that while the TFD purification makes entropy counting easier, it does not actually tell us the properties of the heat bath and the interaction of the ensemble of fuzzballs with it.

\section*{Acknowledgements}

I would like to thank Steve Avery, Bartek Czech, Samir Mathur, Ashoke Sen and Erik Verlinde for helpful discussions.

\bibliographystyle{toine}
\bibliography{Papers}

\end{document}